\documentclass[prd,twocolumn,showpacs,floatfix,amsmath,nofootinbib,amssymb,floatfix]{revtex4}
\usepackage{graphicx,color,dcolumn,booktabs,bm,multirow}
\usepackage{longtable,lscape}
\usepackage{txfonts}
\usepackage{overpic}
\usepackage{amssymb}
\usepackage{indentfirst}
\usepackage{feynmf}   
\usepackage{slashed}  
\usepackage{cases}
\usepackage{color}
\usepackage{multirow}
\usepackage{epstopdf}
\usepackage{graphicx,color,dcolumn,booktabs,bm}
\usepackage{epstopdf}
\usepackage{ulem}
\usepackage[colorlinks, citecolor=green,anchorcolor=red,menucolor=red, linkcolor=blue,filecolor=red,runcolor=red,urlcolor=blue,frenchlinks=red]{hyperref}

\def\lrpartial{\buildrel\leftrightarrow\over\partial}

\usepackage{ulem}

\begin{document}
\title{Possible $B^{(\ast)} \bar{K}$ hadronic molecule state}
\author{Cheng-Jian Xiao$^{1,3}$}
\author{Dian-Yong Chen$^{1,2}$\footnote{Corresponding author}}\email{chendy@impcas.ac.cn}
\affiliation{$^1$Institute of Modern Physics, Chinese Academy of Sciences, Lanzhou 730000, People's Republic of China\\
$^2$Research Center for Hadron and CSR Physics, Lanzhou
University $\&$ Institute of Modern Physics of CAS, Lanzhou 730000,
People's Republic of China\\
$^3$ University of Chinese Academy of Sciences, Beijing 100049, People's Republic of China}

\date{\today}

\begin{abstract}
In the present work, we estimate the decays of the $X(5568)$ and $X(5616)$ in a $B \bar{K}$ and a $B^\ast \bar{K}$ $S$-wave hadronic molecule scenarios, respectively, which may corresponding to the structure observed by D0 Collaboration. Our estimation indicates both $B\bar{K}$ and $B^\ast \bar{K}$ hadronic molecule decay widths could explain the experimental data in a proper model parameter range.
\end{abstract}

\maketitle

\section{Introduction}\label{sec1}
Very recently, the D0 Collaboration observed a new narrow structure in the $B_s^0 \pi^\pm$ invariant mass spectrum, named $X(5568)$, the mass and width of the structure are $5567.8 \pm 2.9 (\mathrm{stat})^{ +0.9}_{ -1.9} (\mathrm{syst})$ MeV and $21.9 \pm 6.4(\mathrm{stat}) ^{+5.0}_{-2.5} (\mathrm{syst})$ MeV, respectively \cite{D0:2016mwd}. The observed channel indicates the isospin of the $X(5568)$ is 1. If the structure decays into $B_s^0 \pi^\pm$ via a $S-$wave, the quantum numbers of the $X(5568)$ are $J^{PC}=0^{++}$.  As indicated in Ref. \cite{D0:2016mwd}, the observed structure may decay through the chain $B_s^{0\ast} \pi^0,\ B_s^{0\ast} \to B_s^{0}\gamma$, in which the soft photon may not be detected since the mass gap of $B_s^{0\ast}$ and $B_s^{0}$ is less than 50 MeV. In this case, this structure decays into $B_s^{\ast 0} \pi^\pm$, and the mass of the structure should be shifted by addition of the mass difference of $B_s^{\ast 0}$ and $B_s^{0}$, which is about $5616$ MeV, while the width remains unchanged. This structure would be named $X(5616)$ with the quantum number $J^{PC}=1^{++}$ if it couples to $B_s^{0\ast} \pi^\pm$ via the lowest $S-$wave.

The quark components of the $B_s^{(\ast)0 }$ and $\pi^\pm$ are $\bar{b}s$ and $u\bar{d}/\bar{u}d$, respectively, thus $X(5568)$ is a kind of structures with four different flavors of quarks, which is observed for the first time. This peculiar property of the $X(5568)$ indicates that it could not be a conventional meson formed by a quark and a antiquark. For a system composed of two quarks and two antiquarks, the color factorization property of the system indicates a two-mesons deuteron-like hadronic molecule structure. Some newly observed hadron states have been extensively investigated in hadronic molecule scenario. For example, in Refs. \cite{Cleven:2013mka, Wang:2013cya, Li:2013yla, Dong:2014zka, Ding:2008gr}, the properties of the $Y(4260)$ were studied, in which the $Y(4260)$ were assigned as a $D D_1(2420)$ molecule. Similarly, the masses and decay behaviors of the $Z_c(3900)/Z_c(4020)$ and $Z_b(10610)/Z_b(10650)$ were estimated in the $D^\ast\bar{D}+h.c/D^\ast \bar{D}^\ast$ and $B^\ast \bar{B}+h.c /B^\ast \bar{B}^\ast$ hadronic molecular pictures, respectively \cite{Sun:2011uh, Aceti:2014uea, Wang:2013daa, Chen:2015ata, Wilbring:2013cha, Dong:2013iqa, Gutsche:2014zda, Chen:2015igx}.

As for the discussed $X(5568)$ or $X(5616)$, we can find that this structure could be decomposed into a bottom-strange meson and a light meson or a bottom meson and a strange meson. For the former case, the bottom-strange meson and light meson could not interact by exchanging a conventional quark-antiquark meson and form a bound state. While for the later case, the bottom and strange mesons could form a hadronic molecule state by exchanging a proper light meson. We notice that the thresholds of $B\bar{K}$ and $B^\ast \bar{K}$ are $5777$ and $5822$ MeV, respectively \cite{Agashe:2014kda}. The observed masses of the $X(5568)$ and $X(5616)$ are about 200 MeV below the thresholds of $B\bar{K}$ and $B^\ast \bar{K}$, respectively. Thus, we can assign the $X(5568)$ and $X(5616)$ reported by the D0 Collaboration as a deeply bounded $B \bar{K}$ or $B^\ast \bar{K}$ hadronic molecule state with $I=1$, which is different with the one discussed in Ref. \cite{Faessler:2008vc}.

To further check the possibilities of the $X(5565)$ and $X(5616)$ as a $B\bar{K}$ and a $B^\ast \bar{K}$ hadronic molecule state, we study the strong decays of the $B\bar{K}$ and $B^\ast \bar{K}$ hadronic molecule states in present work. Due to the kinematic limit, the $X(5565)$ and $X(5616)$ can only strongly decay into $B_s \pi$ and $B_s^\ast \pi$, respectively, which is the observed channel of these states. The approach for describing and treating the deuteron-like hadronic molecule state has been proposed in Refs. \cite{Weinberg:1962hj,Salam:1962ap} and developed in Refs. \cite{Faessler:2007gv, Faessler:2007cu, Branz:2010sh}. This approach has been widely used to study the decay behaviors of the hadronic molecule \cite{Faessler:2007gv, Faessler:2007cu, Branz:2010sh, Dong:2013iqa, Gutsche:2014zda, Dong:2014zka, Xiao:2016hoa, Chen:2016ncs, Chen:2015igx}.

This work is organized as follows: The molecule structure of the $X(5568)$ and $X(5616)$ and their decays are present in the following section. The numerical results and discussions for the decays are presented in Section \ref{sec3}, and Section \ref{sec4} is dedicated to a short summary.

\section{hadronic molecule structure of the $X(5568)$ and $X(5616)$ and their decays}\label{se2}

{\it hadronic molecule structure}.--- In the present work, we consider the $X(5568)$ and $X(5616)$ as hadronic molecules with $I=1$, represented by a $B\bar{K}$ and $B^\ast \bar{K}$ $S-$wave bound state, respectively. The $J^{PC}$ quantum numbers of the $X(5568)$ and $X(5616)$ are $0^{++}$ and $1^{++}$, respectively. The interactions of the hadronic molecule states with their components could be described by the effective Lagrangian, which are
\begin{eqnarray}
\mathcal{L}_{XBK}&=&g_{XBK} X^+(x) \int dy \Phi_X(y^2) B^+(x+\omega_{KB} y) \nonumber\\ &&\times \bar{K}^0(x-\omega_{BK} y),\nonumber\\
\mathcal{L}_{X^\prime B^\ast K} &=&g_{X^\prime B^\ast K} X^{\prime \mu}(x) \int dy  \Phi_{X^\prime}(y^2) B^{\ast +}(x+\omega_{KB^\ast} y)\nonumber\\ &&\times  \bar{K}^0(x-\omega_{B^\ast K} y),
\end{eqnarray}
where $X$ and $X^\prime$ refer to $X(5568)$ and $X(5616)$, respectively. $\omega_{ij}= m_i/(m_i+m_j)$ is the kinematical parameter. The correlation function $\Phi(y)$ is employed to describe the distributions of the molecular components in the hadronic molecule and to render the Feynman diagrams ultraviolet finite as well. In the present calculations, a Gaussian form of the correlation function is adopted, which has been widely used to investigate the hadronic molecule decays \cite{Faessler:2007gv, Faessler:2007cu, Branz:2010sh, Dong:2013iqa, Gutsche:2014zda, Dong:2014zka, Xiao:2016hoa, Chen:2016ncs, Chen:2015igx}. The Fourier transform of the correlation function $\tilde{\Phi}(P_E)$ is in the form,
\begin{eqnarray}
\tilde{\Phi}(P_E) \dot{=} \mathrm{exp}(-P_E^2/ \Lambda^2)
\end{eqnarray}
where $P_E$ is the Jacobi momentum in the Euclidean space, and $\Lambda$ is a model parameter which characterizes the distribution of the components in the hadronic molecule. The concrete value of the $\Lambda$ should be of order 1 GeV and dependent on a different molecular system \cite{Faessler:2007gv, Faessler:2007cu, Branz:2010sh, Dong:2013iqa, Gutsche:2014zda, Dong:2014zka, Xiao:2016hoa, Chen:2016ncs, Chen:2015igx}.

\begin{figure}[t]
\centering
\includegraphics[width=80mm]{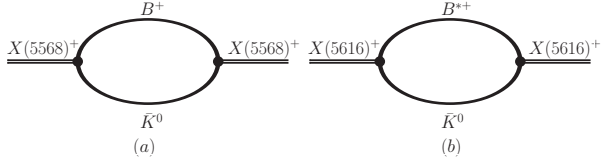}
\caption{The mass operators of the $X(5568)^+$ (diagram (a)) and $X(5616)^+$ (diagram (b)), where $X(5568)^+$ and $X(5616)^+$ are assigned as the $B^+\bar{K}^0$ and $B^{\ast+} \bar{K}^0$ hadronic molecule, respectively. }\label{fig:mo}
\end{figure}

The coupling strength of a hadronic molecule to its components could be evaluated from the compositeness condition \cite{Weinberg:1962hj,Salam:1962ap}, in which the renormalization constant of a composite particle wave function is zero. As for the discussed scalar hadronic molecule, $X(5568)$, the renormalization constants is
\begin{eqnarray}
Z_{X }&\equiv&1-\Sigma^\prime _{X}(m^2_{X})=0, \label{eq:compositeness-con}
\end{eqnarray}
where, $\Sigma^\prime _{X }(m^2_{X})$ is derivative of the mass operator of the $X(5568)$, which is presented in Fig.~\ref{fig:mo}-(a). The concrete form of the mass operator of the $X(5568)$ is
\begin{eqnarray}
\Sigma_{X(5568)} &=& g_{XBK}^2\int \frac{d^4q}{(2\pi)^4}
    \tilde\Phi_X^2[-(q-\omega_{BK}p)^2]\nonumber\\
&&\times\frac{1}{(p-q)^2-m_K^2}\frac{1}{q^2-m_B^2}.
\end{eqnarray}

As a pseudo-vector hadronic molecule, the mass operator of the $X(5616)$ includes the transverse and longitudinal components, which is,
\begin{equation}
\Sigma^{\mu\nu}_{X^\prime}(p)=g^{\mu\nu}_{\perp}
\Sigma_{X^\prime}(p^2)+\frac{p^{\mu}p^{\nu}}{p^2}
\Sigma_{X^\prime}^{L}(p^2),
\end{equation}
with $g^{\mu\nu}_{\perp}=g^{\mu\nu}-p^{\mu}p^{\nu}/p^2$. The $\Sigma_{X^\prime}(p^2)$ and $\Sigma^L_{X^\prime}(p^2)$ are the conventional transverse and longitudinal components, respectively. As shown in Fig.~\ref{fig:mo}-(b), the mass operator of $X(5616)$ is
\begin{eqnarray}
\Sigma_{X(5616)}^{\mu \nu} &=& g_{X^\prime B^\ast K}^2\int \frac{d^4q}{(2\pi)^4}
    \tilde\Phi_{X^\prime}^2[-(q-w_{B^\ast K} p)^2]\nonumber\\
&&\times\frac{1}{(p-q)^2-m_K^2}\frac{-g^{\mu\nu}+q^{\mu}q^{\nu}/m_{B^\ast}^2}
    {q^2-{m_{B^\ast}^2}}.
\end{eqnarray}
The compositeness condition of the $X(5616)$ indicates
\begin{eqnarray}
Z_{X^\prime }&\equiv&1-\Sigma^\prime _{X^\prime}(m^2_{X^\prime})=0. \label{eq:compositeness-con1}
\end{eqnarray}

{\it Decays of the hadronic molecule.}--- We calculate the hadronic molecule decays in an effective Lagrangian approach. The Lagrangians related to the bottom mesons and the light mesons are\cite{Casalbuoni:1996pg,Colangelo:2002mj,Chen:2014sra},
\begin{eqnarray}
\mathcal{L}_{B^{(\ast)}B^{(\ast)}\mathcal{V}}&=&-ig_{BBV}B_i^{\dagger}
   \lrpartial_{\mu}B^j(\mathcal{V}^{\mu})^i_j\nonumber\\
   &&+ig_{B^*B^*V}B_i^{*\nu\dagger}\lrpartial_{\mu}
     B_{\nu}^{*j}(\mathcal{V}^{\mu})^i_j\nonumber\\
    &&+4if_{B^*B^*V}B_{i\mu}^{*\dagger}(\partial^{\mu}\mathcal{V}^{\nu}-\partial^{\nu}
     \mathcal{V}^{\mu})^i_jB_{\nu}^{*j},\\
\mathcal{L}_{B^{(*)}B^{*}\mathcal{P}}&=&-ig_{B^*BP}(\bar B\partial_{\mu}\mathcal{P}B^{*\mu}-{\bar B}^{*\mu}\partial_{\mu}{\mathcal P}B)\nonumber\\
&&+\frac{1}{2}g_{B^*B^*P}\epsilon_{\mu\nu\alpha\beta}\bar B^{*\mu}\partial^{\nu}\mathcal{P}
\lrpartial{}^{\alpha}B^{*\beta},\nonumber
\end{eqnarray}
where $B^{(*)\dagger}=({B}^{(*)-},\bar B^{(*)0},\bar B^{(*)0}_s)$ and $A\lrpartial B=A\partial B-B\partial A$. The $\mathcal{V}$ and $\mathcal{P}$ are the matrixes of the vector nonet and the pseudoscalar nonet, i.e.,
\begin{equation}
\hspace*{\fill}\mathcal{V}=\left(
\begin{matrix}
\frac{1}{\sqrt2}(\rho^0+\omega)&\rho^{+}&K^{*+}\\
\rho^{-}&\frac{1}{\sqrt2}(-\rho^{0}+\omega)&K^{*0}\\
K^{*-}&\bar K^{*0}&\phi
\end{matrix}
\right)\hspace*{\fill},
\end{equation}
\begin{center}
\begin{equation}
\mathcal{P}=
\left(\begin{array}{ccc}
\frac{\pi^{0}}{\sqrt 2}+\frac{\eta_8}{\sqrt6}+\frac{\eta_1}{\sqrt3} &\pi^{+} &K^{+}\\
\pi^{-} &-\frac{\pi^{0}}{\sqrt2}+\frac{\eta_8}{\sqrt6}-\frac{\eta_1}{\sqrt3} &K^{0}\\
K^{-} &\bar K^{0} &-\frac{2\eta_8}{\sqrt6}+\frac{\eta_1}{\sqrt3}
\end{array}\right).
\end{equation}
\end{center}
The interaction between strange mesons and pion  constructed by hidden local gauge symmetry is \cite{Liu:2005jb},
\begin{eqnarray}
\mathcal{L}_{K^\ast K\pi}=-ig_{K^\ast K\pi} K_{\mu}^{\ast \dagger}\vec\pi\cdot\vec\tau\lrpartial{}^{\mu}K,
\end{eqnarray}
where $\vec\tau$ is the Pauli-Dirac matrix and $\vec\pi$ is the pion isospin triplet. The $K$ and $K^\ast $ are the doublets of pseudoscalar and vector strange mesons, respectively.


\begin{figure}[hbt]
\begin{tabular}{cc}
\includegraphics[scale=0.35]{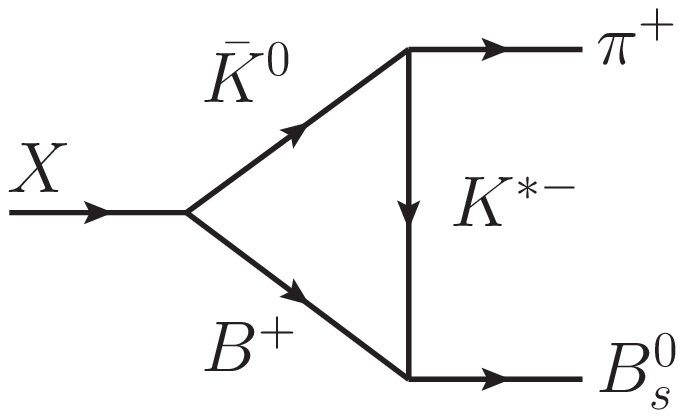}
&\includegraphics[scale=0.35]{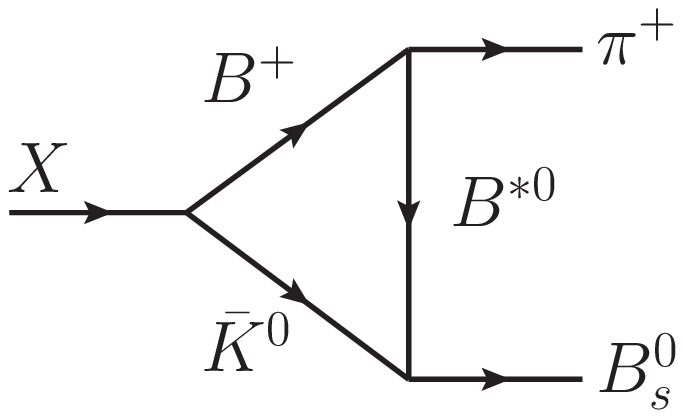}\\
(1) &(2)
\end{tabular}
\caption{Diagrams contributing to process $X(5568) \to B_s^{0}\,\pi^+$.\label{fig:tribspi} }
\end{figure}

As a $B\bar{K}$ hadronic molecule, the $X(5568)$ state can decay to $B_s^0\pi^+$ via S-wave. The amplitudes corresponding to the diagrams in Fig.~\ref{fig:tribspi} are
\begin{eqnarray}
\mathcal{M}_1&=&(i)^3\int \frac{d^4q}{(2\pi)^4}\big[g_{XBK}\tilde\Phi_{X}(-P_{12}^2)\big]\nonumber\\
&&\times\big[-i\sqrt2g_{K^\ast K\pi}(ip_1^\mu+ip_3^\mu)\big]\big[-ig_{BBV}(ip_2^\nu+ip_4^\nu)\big]\nonumber\\
&&\times\frac{1}{p_1^2-m_1^2}
    \frac{1}{p_2^2-m_2^2}
    \frac{-g^{\mu\nu}+q^{\mu}q^{\nu}/m_q^2}{q^2-m_q^2},\nonumber\\
\mathcal{M}_2&=&(i)^3\int \frac{d^4q}{(2\pi)^4}\big[g_{XBK}\tilde\Phi_{X}(-P_{12}^2)\big]\nonumber\\
&&\times\big[ig_{B^\ast BP}(-ip_3^\mu)\big]\big[-ig_{B^\ast BP}(ip_2^\nu)\big]\nonumber\\
&&\times\frac{1}{p_1^2-m_1^2}
    \frac{1}{p_2^2-m_2^2}
    \frac{-g^{\mu\nu}+q^{\mu}q^{\nu}/m_q^2}{q^2-m_q^2},
\end{eqnarray}
where $P_{12}=p_1 \omega_{21}-p_2 \omega_{12}$. The total amplitude of the process $X(5568) \to B_s^0\pi^+$ is
\begin{eqnarray}
&&\mathcal M^{\text{tot}}_{X\to B_s^0\pi^+}=\mathcal M_1+\mathcal M_2,
\end{eqnarray}
and the decay width of $X\to B_s^0\pi^+$ is
\begin{equation}
\Gamma(X(5568) \to B_s^0\pi^+)=\frac{1}{8\pi}\frac{|\vec p|}
    {m_X^2}|{{\mathcal M}}^{\text{tot}}_{X\to B_s^0\pi^+}|^2.
\end{equation}


\begin{figure}[htb]
\begin{tabular}{ccc}
\includegraphics[scale=0.35]{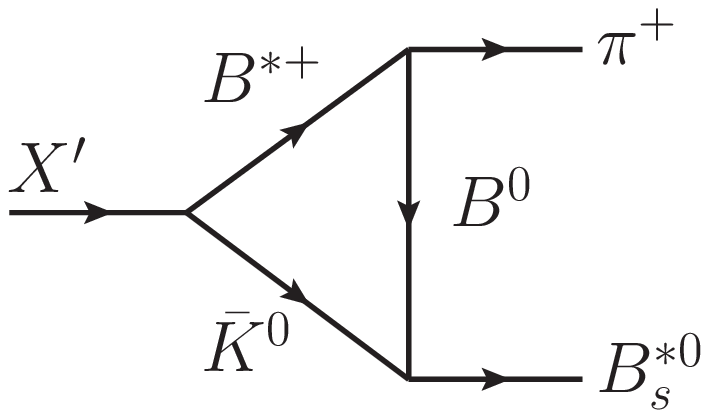}
&\includegraphics[scale=0.35]{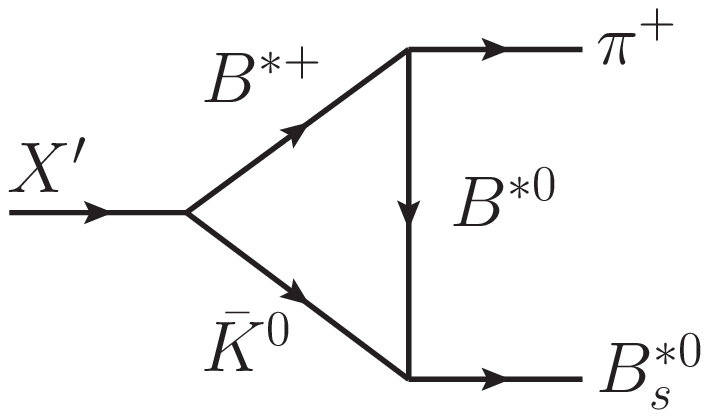}
&\includegraphics[scale=0.35]{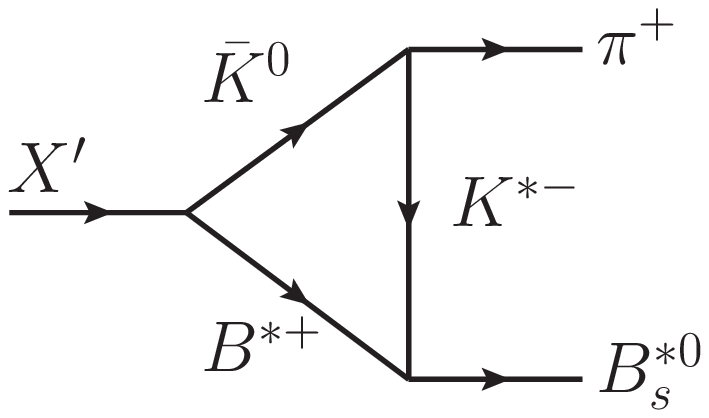}\\
(1) &(2)  &(3)
\end{tabular}
\caption{Diagrams contributing to process $X(5616)\to B_s^{\ast 0}\pi^+$.\label{fig:tri-bsstarpi}}
\end{figure}

The $X(5616)$ state can decay to $B_s^{*0}\pi^+$ via the triangle diagrams presented in Fig.~\ref{fig:tri-bsstarpi}. The corresponding amplitudes are
\begin{eqnarray}
\mathcal{M}_1&=&(i)^3\int \frac{d^4q}{(2\pi)^4}\big[g_{X^\prime BK}\tilde\Phi_{X^\prime}(-P_{12}^2)\big]\nonumber\\
&&\times\big[-ig_{B^\ast BP}(-ip_3^\mu)\big]\big[ig_{B^\ast BP}(ip_2^\nu)\big]\nonumber\\
&&\times\frac{-g^{\phi\mu}+p_1^{\phi}p_1^{\mu}/m_1^2}{p_1^2-m_1^2}
    \frac{1}{p_2^2-m_2^2}
    \frac{1}{q^2-m_q^2},\nonumber\\
\mathcal{M}_2&=&(i)^3\int \frac{d^4q}{(2\pi)^4}\big[g_{X^\prime BK}\tilde\Phi_{X^\prime}(-P_{12}^2)\big]\nonumber\\
&&\times\big[\frac12g_{B^\ast B^\ast P}\epsilon_{\mu\nu\alpha\beta}(-ip_3^\nu)(ip_1^\alpha+iq^\alpha)\big]\nonumber\\
&&\times\big[\frac12g_{B^\ast B^\ast P}\epsilon_{\eta\tau\rho\sigma}(ip_2^\tau)(iq^\rho+ip_4^\rho)\big]\nonumber\\
&&\times\frac{-g^{\phi\beta}+p_1^{\phi}p_1^{\beta}/m_1^2}{p_1^2-m_1^2}
    \frac{1}{p_2^2-m_2^2}
    \frac{-g^{\mu\sigma}+q^{\mu}q^{\sigma}/m_q^2}{q^2-m_q^2},\nonumber\\
\mathcal{M}_3&=&(i)^3\int \frac{d^4q}{(2\pi)^4}\big[g_{X^\prime BK}\tilde\Phi_{X^\prime}(-P_{12}^2)\big]\nonumber\\
&&\times\big[i\sqrt2g_{K^\ast K\pi}(ip_1^\mu+ip_3^\mu)\big]
    \big[ig_{B^\ast B^\ast V}g^{\rho\tau}(ip_2^\eta+ip_4^\eta)\nonumber\\
&&\phantom{\times}+4if_{B^\ast B^\ast V}(iq^\rho g^{\eta\tau}-iq^\tau g^{\eta\rho})\big]\nonumber\\
&&\times\frac{1}{p_1^2-m_1^2}
    \frac{-g^{\phi\tau}+p_2^{\phi}p_2^{\tau}/m_2^2}{p_2^2-m_2^2}
    \frac{-g^{\mu\eta}+q^{\mu}q^{\eta}/m_q^2}{q^2-m_q^2},
\end{eqnarray}
where $P_{12}=p_1\omega_{21}-p_2 \omega_{12}$ and  the total amplitude of process $X(5616)\to B_s^{*0}\pi^+$ is
\begin{eqnarray}
&&\mathcal M^{\text{tot}}_{X^\prime \to B_s^{*0}\pi^+}=\mathcal M_1+\mathcal M_2+\mathcal M_3,
\end{eqnarray}
and the decay width of $X(5616) \to B_s^{\ast 0}\pi^+$
\begin{equation}
\Gamma(X(5616)\to B_s^{\ast 0}\pi^+)=\frac{1}{3}\frac{1}{8\pi}\frac{|\vec p|}
    {m_{X^\prime }^{2}}|{\overline{\mathcal M}}^{\text{tot}}_{X^\prime\to B_s^{\ast 0}\pi^+}|^2.
\end{equation}
where the overline indicates sum over polarizations of vector mesons.

\section{numerical results and discussions}\label{sec3}

The coupling constants of the bottom mesons to the light mesons could be evaluated by heavy quark limit and chiral symmetry \cite{Casalbuoni:1996pg}. the coupling constant $g_{B^{(\ast)}B^\ast P}$ is related to a gauge coupling constant $g$ by
\begin{eqnarray}
g_{B^\ast B^\ast P}=\frac{2g}{f_{\pi}}\,,\quad g_{B^\ast BP}=\frac{2g}{f_{\pi}}\sqrt{m_{B^\ast }m_B}\,,
\end{eqnarray}
where $f_{\pi}=132$ MeV is the pion decay constant and $g=0.44\pm0.03^{+0.01}_{-0.00}$ is determined by
the lattice QCD calculation\cite{Becirevic:2009yb}. The coupling constants concerning to  the bottom mesons and the light vector mesons are
\cite{Lin:1999ad,Oh:2000qr},
\begin{eqnarray}
g_{BBV}=\beta g_V/\sqrt2,\ g_{B^\ast B^\ast V}=\beta g_V/\sqrt2,\ f_{B^\ast B^\ast V}=\lambda m_{B^\ast}g_V/\sqrt2,\nonumber\\
\end{eqnarray}
where the gauge couplings $\beta=0.9$, $\lambda=0.56\,\text{GeV}^{-1}$, and
$g_V=m_\rho/f_\pi$. We use the same $g_{K^\ast K\pi}$ strong coupling
constants as used in Ref.~\cite{Liu:2005jb},
\begin{equation}
g_{K^\ast K\pi}=3.21.
\end{equation}

\begin{figure}[htb]
\includegraphics[scale=0.37]{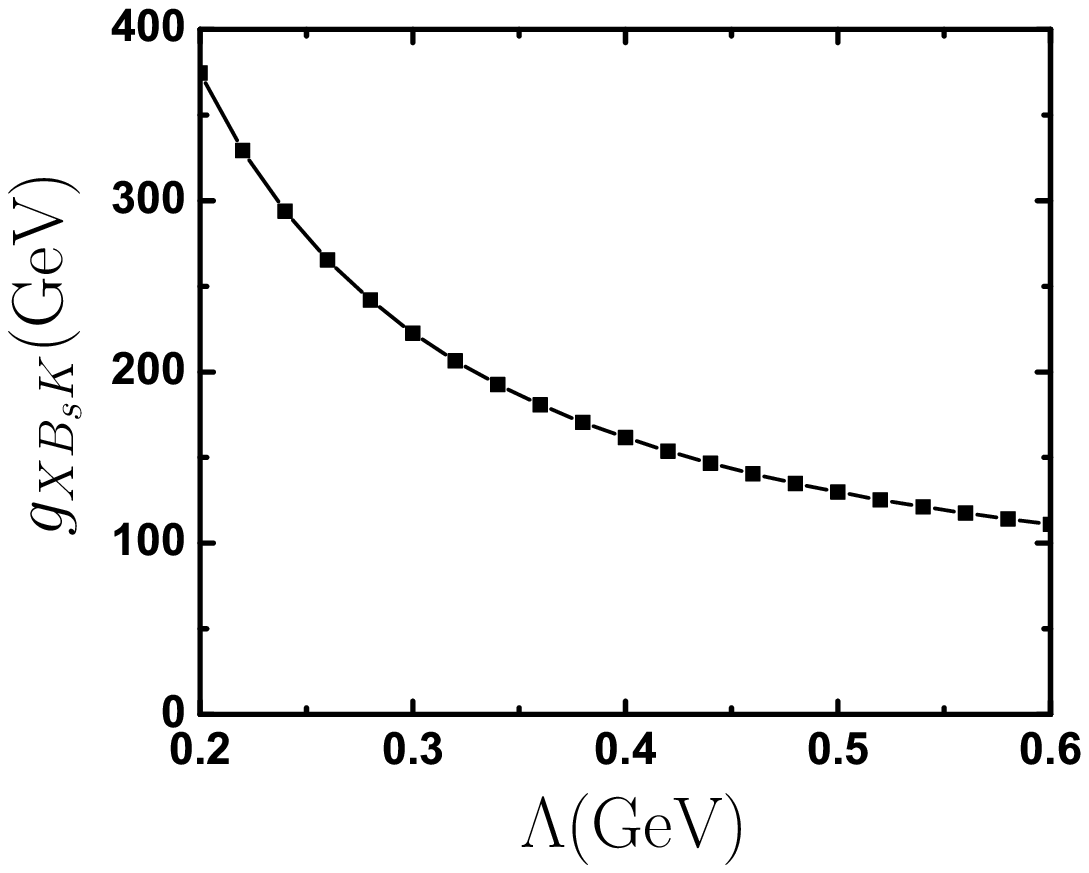}
\includegraphics[scale=0.37]{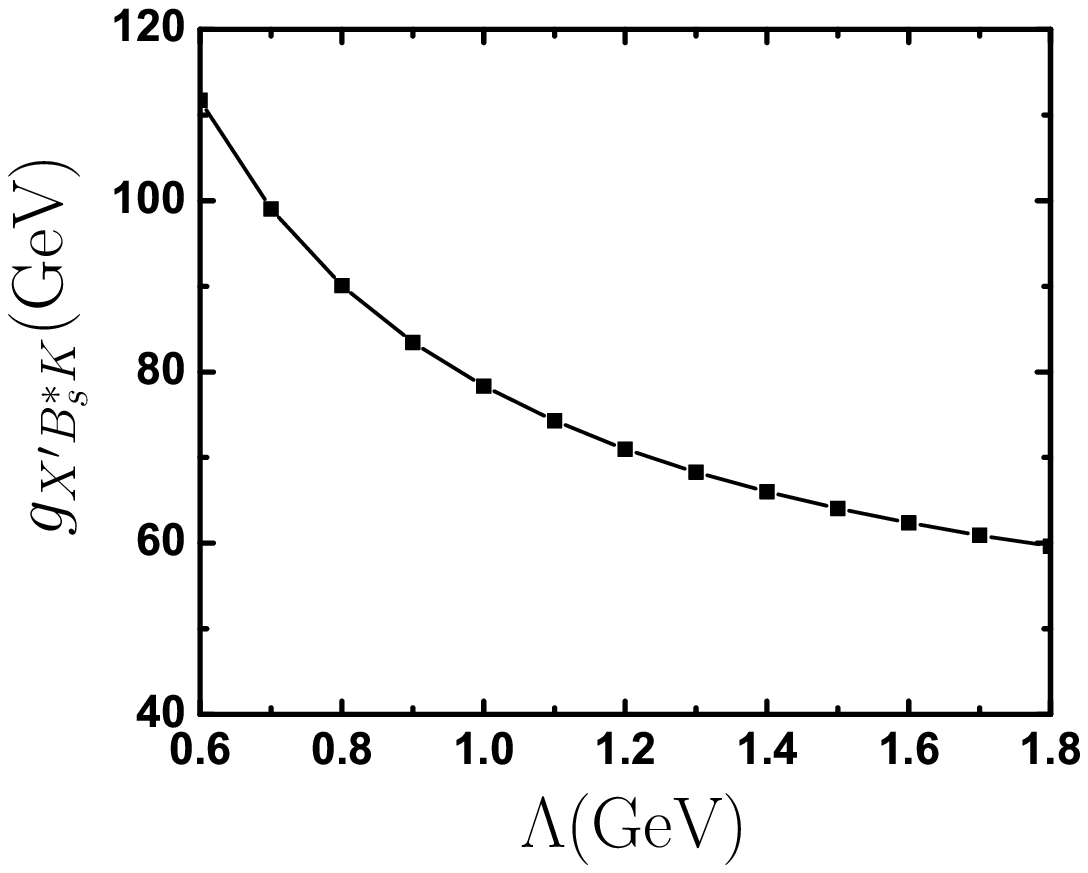}
\caption{The $\Lambda$ dependences of $g_{XB_sK}$ (left column) and $g_{X^\prime B_s^\ast K}$ (right column).\label{fig:couplings}}
\end{figure}

Besides the coupling constants listed above, we estimated the coupling of $g_{XB_s K}$ and $g_{X^\prime B_s^\ast K }$ via compositeness condition presented in Eqs.~(\ref{eq:compositeness-con}) and (\ref{eq:compositeness-con1}). The $\Lambda$ dependences of the coupling constants $g_{XBK}$ and $g_{X^\prime BK}$ are presented in Fig.~\ref{fig:couplings}. Both $g_{XB_sK}$ and  $g_{X^\prime B_s^\ast K}$ decline with the increase of $\Lambda$ in our considered $\Lambda$ range, which is  $0.2 \sim 0.6$ GeV for the $X(5568)$ and $0.6 \sim 1.8$ GeV for the $X(5616)$.

\begin{figure}[htb]
\includegraphics[scale=0.6]{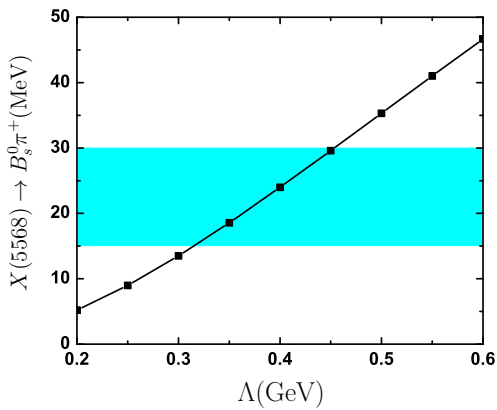}
\caption{Decay width of process $X(5568)\to B_s^0\pi^+$ depending on $\Lambda$, while the blue band is the total width of the $X(5568)$.\label{fig:bspi}}
\end{figure}
\begin{figure}[htb]
\includegraphics[scale=0.6]{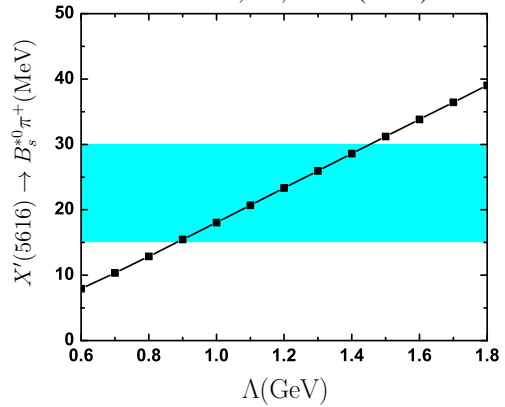}
\caption{Decay width of process $X(5616)\to B_s^{\ast 0}\pi^+$ depending on $\Lambda$. The blue band is total decay width of the $X(5616)$.\label{fig:bsstarpi}}
\end{figure}

The partial width of $X(5568)^+\to B_s^0\pi^+$ is presented in Fig.~\ref{fig:bspi}. It varies from 5.2 to 46.7\,MeV with the variation of $\Lambda$ from 0.2 to 0.6\,GeV. The blue band is the total width of the $X(5568)$ \cite{D0:2016mwd}. The overlapped $\Lambda$ range is $0.31\,\mathrm{GeV} \sim 0.46\,\mathrm{GeV}$. We present the partial decay width of $X(5616)\to B_s^{\ast0}\pi^+$ in Fig.~\ref{fig:bsstarpi}. It varies from 7.9 to 39.0\,MeV with the increasing of $\Lambda$ from 0.6 to 1.8\,GeV. The constrained $\Lambda$ range is $0.89 \sim 1.43\, \mathrm{GeV}$. As shown in Figs.~\ref{fig:bspi} and \ref{fig:bsstarpi}, the partial widths of $X(5568)\to B_s^0\pi^+$ and $X(5616)\to B_s^{\ast0}\pi^+$ depend on the model parameter $\Lambda$. Our calculations indicate that both constrained $\Lambda$ ranges are of order 1.0\,GeV, which are in acceptable range.

There is another possibility that the experimental $X(5568)$ structure contains two states, i.e, the $X(5568)$ and $X(5616)$, which would overlap in the $B_s \pi $ invariant mass spectra. In this case, the present experimental measurement could not distinguish this two states \cite{D0:2016mwd}.  However, the $X(5616)$ state can radiatively transit into $B_s^\ast$ and $B_s$, while the $X(5568)$ can only decay to $B_s^\ast\gamma$. The future experimental measurements of the radiative decays of the neutral $X(5568)$ and $X(5616)$ could provide us more information about the structure observed by the D0 Collaboration \cite{D0:2016mwd}.

\section{summary}\label{sec4}
To summarize, we interpret the newly observed structure $X(5568)$ and $X(5616)$ as $S-$wave deeply bound state of $B\bar{K}$ and $B^\ast \bar{K}$, respectively. With an effective Lagrangian  approach, we estimate the partial decay widths $X(5568)\to B_s^0\pi^+$ and $X(5616)\to B_s^{\ast0}\pi^+$. Our results indicate both explanations could be acceptable since the constrained model parameter $\Lambda$ are of order 1.0\,GeV. Since both $BK$ and $B^\ast K$ hadronic molecule could exist in the structure observed by D0 collaboration \cite{D0:2016mwd}, we propose to study these states by radiative decay experimentally, which would help us to further understand the observed structure in the $B_s \pi$ invariant mass spectrum by the D0 Collaboration \cite{D0:2016mwd}.

\section{Acknowledgements}
This project is supported by the National Natural Science Foundation of China under Grant No. 11375240

\end{document}